\documentclass{elsart}

\usepackage{graphicx}
\usepackage{amssymb}

\begin{document}

\begin{frontmatter}

% Title, authors and addresses

% use the thanksref command within \title, \author or \address for footnotes;
% use the corauthref command within \author for corresponding author footnotes;
% use the ead command for the email address,
% and the form \ead[url] for the home page:
% \title{Title\thanksref{label1}}
% \thanks[label1]{}
% \author{Name\corauthref{cor1}\thanksref{label2}}
% \ead{email address}
% \ead[url]{home page}
% \thanks[label2]{}
% \corauth[cor1]{}
% \address{Address\thanksref{label3}}
% \thanks[label3]{}

\title{On the distribution of 1550-nm photon pairs efficiently generated using
a periodically poled lithium niobate waveguide}

% use optional labels to link authors explicitly to addresses:
% \author[label1,label2]{}
% \address[label1]{}
% \address[label2]{}

\author{Shigehiko Mori, Jonas S\"{o}derholm, Naoto Namekata, Shuichiro Inoue}

\address{Institute of Quantum Science, Nihon University, 1-8 Kanda-Surugadai, Chiyoda-ku, Tokyo 101-8308, Japan}

\begin{abstract}
We report on the generation of photon pairs in the 1550-nm band
suitable for long-distance fiber-optic quantum key distribution.
The photon pairs were generated in a periodically poled lithium
niobate waveguide with a high conversion-efficiency. Using a
pulsed semiconductor laser with a pulse rate of 800~kHz and a
maximum average pump power of 50~$\mu$W, we obtained a coincidence
rate of 600~s$^{-1}$. Our measurements are in agreement with a
Poissonian photon-pair distribution, as is expected from a
comparison of the coherence time of the pump and of the detected
photons. An average of 0.9 photon pairs per pulse was obtained.
\end{abstract}

\end{frontmatter}

\section{Introduction}

Efficient generation of photon pairs in the 1550-nm fiber-optic
communication band would be useful for practical realizations of
quantum communication, for example, long-distance fiber-optic
quantum key distribution (QKD) \cite{Bennett,Gisin}. Photon pairs
are usually generated at visible wavelengths using the process of
spontaneous parametric down conversion (SPDC) in second-order
nonlinear bulk crystals \cite{Kwiat}. Recently, however,
photon-pair generation in the 1550-nm band has been demonstrated
\cite{Fiorentino,Bonfrate}. In these experiments, large-sized
lasers with high power, such as Ti:sapphire lasers, were used due
to the low conversion efficiency of the photon-pair sources.
However, a periodically poled lithium niobate waveguide (PPLN-WG)
has the potential to generate photon pairs much more efficiently
than the bulk crystals previously used. This stems from the fact
that the largest nonlinear coefficient $d_{33}$ can be utilized
through quasi-phase-matching (QPM), and that the guiding structure
permits confinement of the pump beam over the entire interaction
length. Therefore, small, handy, and low-cost semiconductor lasers
can be used in combination with these nonlinear waveguides
\cite{Tanzilli,Sanaka,YoshizawaEL,YoshizawaJJAP}.

In most quantum information applications, the probability for
generating simultaneous multiple photon pairs must be kept low. It
is therefore important to investigate the photon-pair distribution
when characterizing a source. Direct measurements of photon-number
distributions have recently been demonstrated using a
visible-light photon counter \cite{Waks} and a superconducting
transition-edge sensor \cite{Miller}. However, these measurements
were carried out under cryogenic conditions and are too extreme to
allow for common use.

In this paper, we report on efficient generation of photon pairs
at 1550~nm using a PPLN-WG pumped by a pulsed semiconductor laser.
By analyzing the detected single and coincidence counts as
functions of the pump power, we find that our measurements are in
agreement with a Poissonian photon-pair distribution. This is also
found to be in agreement with a comparison of the coherence times
of the pump and photon pairs.

\section{The experimental setup}

\begin{figure}
\centering
\includegraphics[width=9cm]{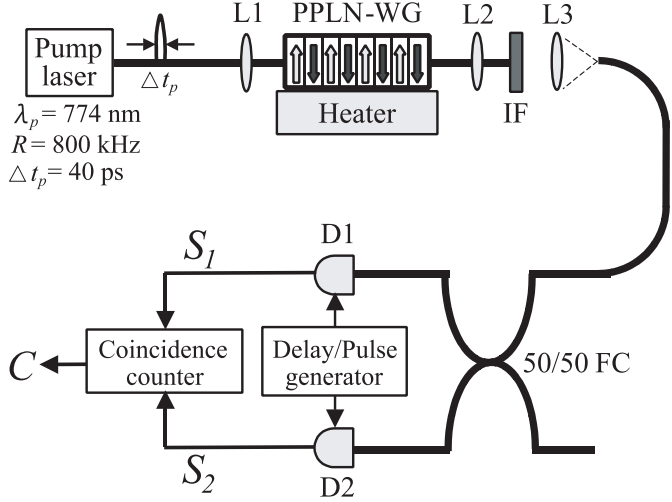}

\caption{Schematic experimental setup. L1, L2, and L3: lenses, IF:
interference filter, 50/50~FC: 50/50 single-mode fiber coupler, D1
and D2: single-photon detectors, $S_1$: count rate at D1, $S_2$:
count rate at D2, $C$: count rate at the coincidence counter.
\label{fig:setup}}
\end{figure}

In Fig.~\ref{fig:setup}, our experimental setup is schematically
depicted. A 3~cm long PPLN-WG (HC Photonics) is pumped by a pulsed
semiconductor laser (Pico Quant PDL 800). The laser light has a
wavelength of 774~nm and is generated in $\sim$40~ps long pulses
with a peak power of $\sim$340~mW and a repetition rate of
800~kHz. QPM for degenerate down-conversion at 1548~nm is obtained
by heating the waveguide to 70$^\circ$C, which also suppresses the
photorefractive effect \cite{Tanzilli}. Both the signal and idler
photon of each photon pair have the same polarization as the pump.
In order to measure only the down-converted photons, the emerging
light from the waveguide goes through an interference filter (IF)
with a full width at half maximum (FWHM) of 30~nm centered around
1550~nm. The transmitted down-converted photons are subsequently
coupled to a 50/50 single-mode fiber coupler (50/50 FC) before
they are detected by two single-photon detectors (D1 and D2). The
detectors are InGaAs/InP avalanche photodiodes (Epitaxx EPM239BA)
operated in gated passive quenching mode at -40$^\circ$C
\cite{NamekataOL}. A delay/pulse generator (Stanford Research
Systems DG535) is used to gate the detectors. The gate pulses are
$\sim$1~ns long and the gating rate is 800~kHz, that is, the same
as the optical pulse rate. The output signals from the detectors
are finally sent to a coincidence counter (Stanford Research
Systems SR400). The maximum value of the coincidence counts is
obtained by synchronizing the gate pulses with the optical pulses
at each detector. The quantum efficiency of D1 was 25\% and its
dark-count probability was $6 \times 10^{-5}$. The corresponding
values for D2 were 10\% and $4 \times 10^{-4}$, respectively.

\section{Photon-pair distributions}

Since our detectors cannot distinguish if one or more photons make
them click, the single-count rate at detector $k$ (when neglecting
dark counts) can be expressed as
\begin{equation}
S_k = R \sum_{m=0}^\infty \frac{p (m)}{2^{2 m}} \sum_{n=0}^{2 m}
\left( \matrix{2 m \cr n} \right) [1 - (1 - T \eta_k)^n] ,
\label{eq:Sk}
\end{equation}
where $R$ is the pulse rate, $p (m)$ denotes the probability for
$m$ photon pairs to be generated, $T$ is the transmittivity of the
optical components, and $\eta_k$ is the detector efficiency.
Similarly, the coincidence-count rate can be written as
\begin{equation}
C = R \sum_{m=0}^\infty \frac{p (m)}{2^{2 m}} \sum_{n=0}^{2 m}
\left( \matrix{2 m \cr n} \right) [1 - (1 - T \eta_1)^n] [1 - (1 -
T \eta_2)^{2 m - n}] . \label{eq:C}
\end{equation}
As we will describe below, the photon-pair distribution $p (m)$
resulting from the SPDC process depends on the experimental
conditions.

\subsection{The degenerate SPDC process}

It is well known that the degenerate SPDC process produces the
squeezed vacuum state when no initial photons are present in the
down-conversion mode. The corresponding photon-pair distribution
\cite{Yuen} can be expressed as
\begin{equation}
p_\mathrm{sv} (\mu,m) = \frac{(2 m - 1)!! \mu^m}{(2 m)!! (\mu +
1)^{m+1/2}} , \label{eq:SqVacDistr}
\end{equation}
where $\mu = \sinh^2 r$ is the average number of photons, and the
squeezing parameter $r$ is proportional to the electric field of
the pump \cite{Scully}. Assuming that the pump pulse has the same
form for all pump powers, we thus obtain
\begin{equation}
\mu = \sinh^2 \sqrt{K P_\mathrm{ave}} , \label{eq:muSqVac}
\end{equation}
where $K$ is a constant and $P_\mathrm{ave}$ is the average pump
power. We note that $\mu \approx K P_\mathrm{ave}$ for small
squeezing parameters $r = (K P_\mathrm{ave})^{1/2} \ll 1$.

\subsection{The nondegenerate SPDC process}

With no initial photons in the signal and idler modes, the
nondegenerate SPDC process produces the two-mode squeezed vacuum
state. The photon-pair distribution is then thermal \cite{Mollow}
\begin{equation}
p_\mathrm{th} (\mu,m) = \frac{\mu^m}{(\mu + 1)^{m+1}} ,
\label{eq:ThermDistr}
\end{equation}
where $\mu$ denotes the average number of photon pairs. Analogous
to the degenerate case, we obtain relation (\ref{eq:muSqVac}).

\subsection{Many distinguishable SPDC processes}

If we instead assume that the detected photons originate from many
distinguishable down-conversion processes, the photon-pair
distribution can be approximated by the Poissonian distribution
\begin{equation}
p_\mathrm{poi} (\nu,m) = \frac{\nu^m e^{-\nu}}{m!} ,
\label{eq:PoiDistr}
\end{equation}
where $\nu$ is the average value of the total number of photon
pairs generated by a single pump pulse \cite{deRiedmatten}. As
there are many distinguishable processes, the average photon-pair
number in each of them is usually small ($\mu \ll 1$), and
therefore proportional to average pump power. For both degenerate
and nondegenerate SPDC, we then get $\nu \approx \mathcal{K}
P_\mathrm{ave}$, where $\mathcal{K}$ is a constant.

\subsection{Experimental results and curve fitting}

The effects of losses are slightly involved due to the fact that
photon pairs are generated throughout the waveguide. The optical
pump power in the waveguide should be well described by $P (x) =
P_0 \exp (-L_\mathrm{p} x)$, where $P_0$ is the optical power
coupled into the guided mode, and $L_\mathrm{p}$ and $x$ are the
loss and propagation distance in the waveguide, respectively. As
discussed above, the average number of generated photon pairs is
proportional to the pump power when $\mu \ll 1$. Denoting the
corresponding constant as $\kappa$ and the length of the waveguide
as $d$, these assumptions give us the following expression for the
average number of photons generated by a single pump pulse
\begin{equation}
N_\mathrm{gen} = 2 \kappa P_0 \int_0^d e^{- L_\mathrm{p} x} \;
\textrm{d} x = \frac{2 \kappa P_0 \left( 1 - e^{- L_\mathrm{p} d}
\right)}{L_\mathrm{p}} .
\end{equation}
Losses for the down-converted light will split photon pairs and
make the photon distribution consist of both even and odd numbers
of photons. Using the notation $L_\mathrm{dc}$ for this loss in
the waveguide, the average number of photons reaching the back
facet at $x = d$ is found to be
\begin{equation}
N_\mathrm{facet} = 2 \kappa P_0 \int_0^d e^{- L_\mathrm{p} x -
L_\mathrm{dc} (d - x)} \; \textrm{d} x = \frac{2 \kappa P_0 \left(
e^{- L_\mathrm{p} d} - e^{- L_\mathrm{dc} d} \right)}
{L_\mathrm{dc} - L_\mathrm{p}} ,
\end{equation}
where we have assumed $L_\mathrm{dc} \neq L_\mathrm{p}$. The
length of our waveguide is $d = 3$~cm and its losses are
approximately $\mathcal{L}_\mathrm{p} = 0.7$~dB/cm and
$\mathcal{L}_\mathrm{dc} = 0.35$~dB/cm, according to the
manufacturer. Since the losses expressed in decibel are related to
the losses in the equations above according to $10 L = \mathcal{L}
\ln 10$, we find the effective internal transmittivity to be
$T_\mathrm{int} = N_\mathrm{facet}/N_\mathrm{gen} \approx 0.880$.
The facet is antireflection coated for 1550~nm, and the
corresponding waveguide-air transmittivity is $T_\mathrm{AR}
\approx 0.99$.

In an independent measurement, using a laser with a wavelength of
1550~nm (Hamamatsu PLP-01), the total transmittivity of the
optical components after the waveguide was found to be
$T_\mathrm{ext} \approx 0.17$. The obtained overall transmittivity
for the down-converted photons in our setup is thus $T =
T_\mathrm{int} T_\mathrm{AR} T_\mathrm{ext} \approx 0.148$.

Using the experimental parameters $R$ = 800 000, $T = 0.148$,
$\eta_1 = 0.25$, and $\eta_2 = 0.10$ together with any of the
photon-pair distributions (\ref{eq:SqVacDistr}),
(\ref{eq:ThermDistr}), or (\ref{eq:PoiDistr}), the corresponding
theoretical curves given by Eqs.~(\ref{eq:Sk}) and (\ref{eq:C})
can be fitted to the experimental data by varying the unknown
parameter $K$ or $\mathcal{K}$. We have used relative weighting
for determining the best curve fits. Hence, we have minimized
expressions of the form $\sum_k (\mathcal{E}_k -
\mathcal{T}_k)^2/\mathcal{E}_k^2$, where $\mathcal{E}_k$ and
$\mathcal{T}_k$ denote the experimental and theoretical value,
respectively, for a given pump power. In this way, the
single-count and coincidence-count errors can be treated
simultaneously, even though their values are very different. For
single degenerate and nondegenerate processes, we obtain the
values $K = 23.2$~mW$^{-1}$ and $K = 14.4$~mW$^{-1}$,
respectively. Assuming a Poissonian photon-pair distribution, the
errors are minimized for $\mathcal{K} = 18.5$~mW$^{-1}$.

\begin{figure}
\centering
\includegraphics[width=13cm]{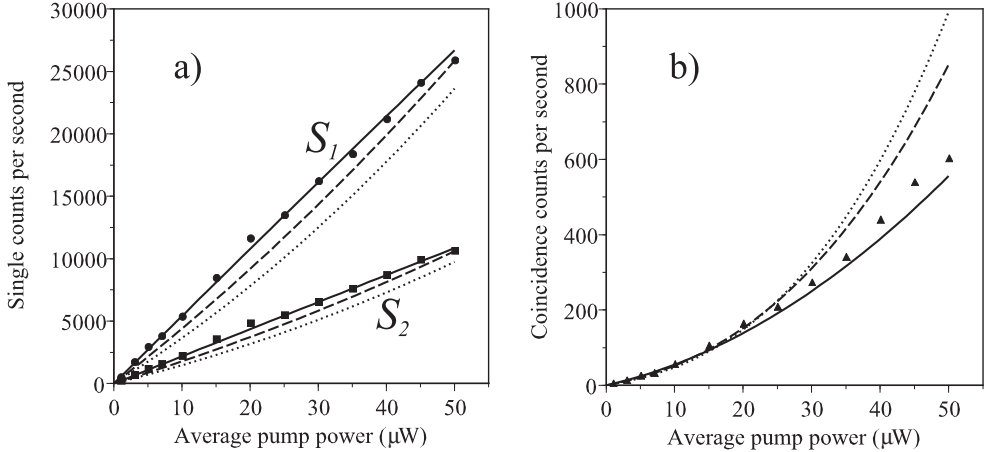}

\caption{Experimental and theoretical count rates. In a), the
circles and squares correspond to the measured single-count rates
at the two detectors for different pump powers. In b), the
triangles represent the measured coincidence-count rates. The dark
counts have been subtracted from the experimental data. The
dotted, dashed, and solid lines are the theoretical curves for a
single degenerate SPDC process, a single nondegenerate SPDC
process, and many distinguishable SPDC processes, respectively.
Hence, the dashed lines correspond to a thermal photon-pair
distribution, and the solid lines to a Poissonian one.
\label{fig:Counts}}
\end{figure}

The experimental results and the curve fits are plotted in
Fig.~\ref{fig:Counts}. Since the measured count rates were far
below saturation of the detectors, we have simply subtracted the
dark-count rates $\delta_1 = 48$~s$^{-1}$ and $\delta_2 =
320$~s$^{-1}$ from the raw single-count data to obtain $S_1$ and
$S_2$, respectively. By measuring the coincidence counts when one
of the two beams after the beam splitter was blocked, the
dark-count-induced coincidence rate $(S_1 \delta_2 + S_2 \delta_1
+ \delta_1 \delta_2)/R$ was experimentally verified. Subtracting
it from the raw coincidence data thus gave us the true
coincidence-count rate $C$ presented in Fig.~\ref{fig:Counts}. The
Poissonian photon-pair distribution is seen to fit well to the
experimental data. The best curve fits for a single degenerate and
a single nondegenerate SPDC process are considerably worse, the
former in particular. The reasons for these differences will be
discussed below.

\subsection{Spectra and coherence times}

The Poissonian photon-pair distribution can be theoretically
justified by comparing the coherence time of the pump
$\tau_\mathrm{p}$ with that of the down-converted light
$\tau_\mathrm{dc}$, as was recently discussed by de Riedmatten
\textit{et al.} \cite{deRiedmatten}. We expect the coherence
length of the laser to be close to the pulse length, that is,
$\tau_\mathrm{p} \approx 40$~ps. The generated photon pairs on the
other hand, have a wide spectrum and the coherence length of the
detected photons should therefore be determined by the 30-nm
filter. We here follow the reasoning in Ref.~\cite{deRiedmatten},
but offer some more detailed arguments. Let us assume that the
spectrum of the light transmitted through the filter can be
approximated with a Gaussian function, whose FWHM is given by the
corresponding wavelengths of the filter. For a general filter,
these wavelengths are given by $\lambda_{1,2} = \lambda_\mathrm{c}
\pm B/2$, where $\lambda_\mathrm{c}$ and $B$ are the filter's
central wavelength and FWHM bandwidth, respectively. Expressed in
the angular frequency $\omega = 2 \pi f$, the spectrum is thus
assumed to satisfy
\begin{equation}
S (\omega) \propto \exp \left\{ - \frac{[\omega - \pi (f_1 +
f_2)]^2 \ln 2}{\pi^2 (f_1 - f_2)^2} \right\} ,
\end{equation}
where $f_k = c/\lambda_k$, $k = 1,2$, and $c$ is the speed of
light in vacuum. Assuming that the down-converted light pulses are
transform limited, the temporal FWHM of the pulse's intensity
$W_t$, and the FWHM of the spectrum $W_f$, are related according
to $W_t W_f = (2 \ln 2)/\pi$. Approximating the coherence length
of the pulses with the FWHM of their intensity, we thus arrive at
\begin{equation}
\tau_\mathrm{dc} = \frac{(4 \lambda_\mathrm{c}^2 - B^2) \ln 2}{2
\pi c B} .
\end{equation}
Usually $B \ll \lambda_\mathrm{c}$, in which case we recover the
relation $\tau_\mathrm{dc} \approx 0.44 \lambda_\mathrm{c}^2/c B$
given in Ref.~\cite{deRiedmatten}.

\begin{figure}
\centering
\includegraphics[width=10.cm]{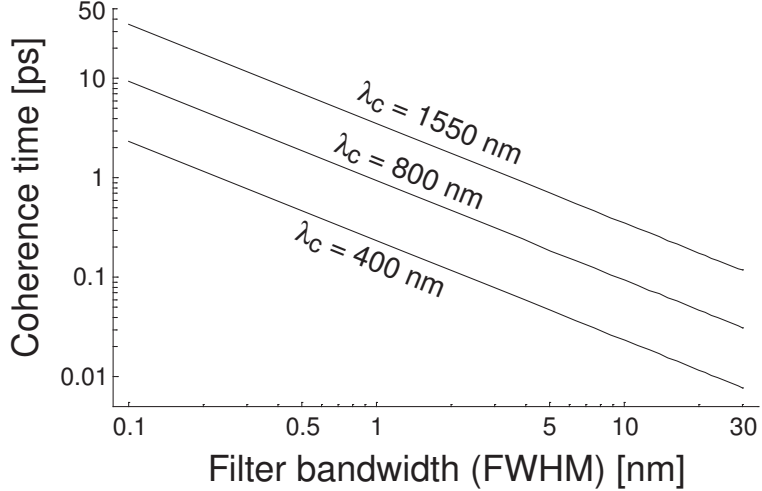}

\caption{The coherence time set by the filter bandwidth when the
transmitted spectrum is assumed to be Gaussian. The three curves
correspond to filters with central wavelengths of 400~nm, 800~nm,
and 1550~nm, respectively. \label{fig:CoherenceTime}}
\end{figure}

In Fig.~\ref{fig:CoherenceTime}, we have plotted the coherence
time as a function of the filter bandwidth for different central
wavelengths of the filter. As $\lambda_\mathrm{c} = 1548$~nm and
$B = 30$~nm for our filter, the coherence time is found to be
$\tau_\mathrm{dc} \approx 118$~fs. We thus have $\tau_\mathrm{p}
\gg \tau_\mathrm{dc}$, which results in a Poissonian photon-pair
distribution \cite{deRiedmatten}. We also note that since the
detected down-converted light has a spectral width of 30~nm, which
is considerably wider than that of the pump, the observed process
cannot be degenerate. In light of this, it appears natural that a
single nondegenerate process gives a better curve fit than a
single degenerate one, as found in Fig.~\ref{fig:Counts}.

According to the plot in Fig.~\ref{fig:CoherenceTime}, we can
obtain $\tau_\mathrm{p} \approx \tau_\mathrm{dc}$ with the present
source, if a very narrow filter with a bandwidth of about one
tenth of a nanometer is used. We would then expect the photon-pair
distribution to be given by a single degenerate or nondegenerate
SPDC process, which would increase the ratio between coincidence
and single counts. However, due to the use of such a narrow
filter, the count rates would then be much smaller. Also, even if
the filter is centered around the frequency for degenerate
down-conversion, the photon distribution may be affected by the
spectral width of the pump in this case.

\section{Comparisons and improvements}

In Table~\ref{tab:comparison}, our experimental results obtained
at the maximum average pump power of 50~$\mu$W are presented.
Here, $S_\mathrm{ave} = (S_1 + S_2)/2$ is the average single-count
rate for the two detectors, and $P_\mathrm{peak}$ denotes the peak
pump power. For comparison, results of previous studies carried
out at 1550~nm are also listed. Semiconductor and Ti:sapphire pump
lasers have been abbreviated as SC and Ti:S, respectively. The
photon-pair sources used in the earlier experiments were: a
PPLN-WG \cite{YoshizawaJJAP}, four wave mixing (FWM) in a
dispersion-shifted fiber (DSF) \cite{Fiorentino}, a periodically
poled silica fiber (PPSF) \cite{Bonfrate}, and a PPLN bulk crystal
\cite{Mori}. Since the contribution to the maximum
coincidence-count rate from independent photon pairs and dark
counts is unknown in the experiments with a cw pump
\cite{Bonfrate,YoshizawaJJAP}, the accidental coincidence-count
rate (measured with a long delay between the two detector signals)
has been subtracted to obtain the values of $C$ in these cases. As
a simple measure of the efficiency of the different setups, we
have calculated the probability of a coincidence count divided by
the average pump power and the repetition rate in the bottom line
of Table~\ref{tab:comparison}. Due to the high conversion
efficiency of the PPLN-WG and the high peak intensity of the
pulsed laser, the highest value is obtained in the present
experiment. However, the coincidence rate is low for a practical
QKD system. Assuming an average of $\nu = 0.1$ generated photon
pairs per pulse, which is the value most frequently adopted in QKD
experiments, and obtained at a pump power of $P_\mathrm{ave} =
5.4$~$\mu$W in the present experiment, the resulting coincidence
rate is only $C = 26$~s$^{-1}$.

\begin{table}
\caption{Comparison of the present experiment and previous studies
in the 1550-nm band. \label{tab:comparison}}

\begin{tabular}{cccccc}
\hline Source & PPLN-WG & PPLN-WG \cite{YoshizawaJJAP} &  FWM in
DSF \cite{Fiorentino} & PPSF \cite{Bonfrate} & Bulk PPLN \cite{Mori} \\
\hline Pump & SC (pulsed) & SC (cw) & Ti:S (pulsed) & Ti:S (cw) & SC (pulsed) \\
$P_\mathrm{ave}$ & 0.05 mW & 0.15 mW & 2 mW & 600 mW & 0.05 mW \\
$P_\mathrm{peak}$ & 0.34 W & NA & 9 W & NA & 0.34 W \\
$R$ & 0.8 MHz & 2 MHz & 0.588 MHz & NA & 0.8 MHz \\
$\eta_1, \eta_2$ & 25\%, 10\% & 19.0\%, 17.3\% & 25\%, 20\% & 1.7\%, 1.4\% & 25\%, 25\% \\
$S_\mathrm{ave}$ & 17 000 s$^{-1}$ & 24 000 s$^{-1}$ & 18 000
s$^{-1}$ & 275 000 s$^{-1}$ $^\mathrm{a}$ & 88 s$^{-1}$ \\
$C$ & 600 s$^{-1}$ & 800 s$^{-1}$ $^\mathrm{b}$ & 1000 s$^{-1}$ & 500 s$^{-1}$ $^\mathrm{b}$ & 0.5 s$^{-1}$ \\
$C/P_\mathrm{ave} R$ & 15 W$^{-1}$ & 2.7 W$^{-1}$ & 0.85 W$^{-1}$ & NA & 0.012 W$^{-1}$ \\
\hline
\multicolumn{4}{l}{$^\mathrm{a}$ Reported in Ref.~\cite{YoshizawaJJAP}.} \\
\multicolumn{4}{l}{$^\mathrm{b}$ Accidental coincidence-count rate
subtracted.}
\end{tabular}

\end{table}

There are several ways to increase the useful rate of coincidences
with current technology. First of all, detector D2 has an
efficiency that is considerable lower than that of D1. It is, of
course, possible to have two detectors with the higher efficiency.

Secondly, the repetition rate can be increased. In our experiment,
we were limited to 800~kHz by the delay/pulse generator. For
repetition rates exceeding 1~MHz, the effects of afterpulsing have
to be considered \cite{NamekataOL}, but a repetition rate of
10~MHz with negligible afterpulse probability has recently been
achieved using discharge-pulse counting \cite{YoshizawaDischarge}.

Thirdly, it is possible to improve the coupling of the
down-converted light into the single-mode fiber \cite{Banaszek}.
In this first experiment, the optics was not optimized and
resulted in an estimated fiber-coupling efficiency of only 25\%.

Fourthly, the useful coincidence rate can be increased by using a
QPM device that generates nondegenerate photon pairs, whose two
photons are of easily separable frequencies
\cite{NamekataPreparation}. As our photon pairs are close to
degenerate, we have to perform beam splitting in order to
simultaneously detect the two constituting photons. This reduces
the number of coincidence counts by half compared to the
distinctly nondegenerate case, in which the photons can be
separated efficiently by a prism, dichroic mirror, or wavelength
division multiplexing module. Moreover, with an appropriate QPM
device, the signal and idler photons can have very different
wavelengths. If the signal wavelength is short enough, a highly
efficient Si-based detector with low dark count can be used to
detect the signal photons, and thereby the signal-to-noise ratio
can be improved drastically \cite{Mason,Pelton,Fasel}.

Using InGaAs/InP detectors and the improvements above, the values
$\eta_2 = 0.25$, $R = 10$~MHz, and $T_\mathrm{ext} = 0.50$ are
feasible. If we further assume that no beam splitting is
necessary, the coincidence rate becomes $C \approx$ 14
000~s$^{-1}$ as the average number of generated photon pairs is
$\nu = 0.1$.

\section{Conclusions}

We have shown that a PPLN-WG is a highly efficient source of
photon pairs at 1550~nm. In the present experiment, the
coincidence rate was found to be too low for a practical QKD
system, but we have argued that a reasonable rate is within reach
of present technology. With these improvements the PPLN-WG should
be a suitable source for long-distance QKD system, since it can be
pumped by a cheap semiconductor laser, and light at the
single-photon level can be generated in the minimum-loss window of
common optical fibers. Our measurements also suggest that the
down-converted light can be described by a Poissonian photon-pair
distribution, which is in accordance with theory. Using the value
$\mathcal{K} = 18.5$ mW$^{-1}$ obtained through curve fitting, we
can deduce that an average number of photon pairs per pulse of
$\nu \approx \mathcal{K} P_\mathrm{ave} = 0.9$ was achieved with a
pump power of $P_\mathrm{ave} = 50$ $\mu$W.

\ack

This work was partly supported by the National Institute of
Information and Communications Technology.


\begin{thebibliography}{00}

\bibitem{Bennett} C.H. Bennett, G. Brassard, Quantum cryptography: Public key
distribution and coin tossing, in: Proceedings of IEEE
International Conference on Computers, Systems, and Signal
Processing, Institute of Electrical and Electronics Engineers, New
York, 1984, pp. 175-179.

\bibitem{Gisin} N. Gisin, G. Ribordy, W. Tittel, H. Zbinden, Rev. Mod. Phys. 74 (2002) 145-195.

\bibitem{Kwiat} P.G. Kwiat, K. Mattle, H. Weinfurter, A. Zeilinger, A.V.
Sergienko, Y. Shih, Phys. Rev. Lett. 75 (1995) 4337-4341.

\bibitem{Fiorentino} M. Fiorentino, P.L. Voss, J.E. Sharping, P. Kumar,
IEEE Photonics Technol. Lett. 14 (2002) 983-985.

\bibitem{Bonfrate} G. Bonfrate, V. Pruneri, P.G. Kazansky,
P. Tapster, J.G. Rarity, Appl. Phys. Lett. 75 (1999) 2356-2358.

\bibitem{Tanzilli} S. Tanzilli, H. De Riedmatten, W. Tittel, H. Zbinden,
P. Baldi, M. De Micheli, D.B. Ostrowsky, N. Gisin, Electron. Lett.
37 (2001) 26-28.

\bibitem{Sanaka} K. Sanaka, K. Kawahara, T. Kuga, Phys. Rev. Lett. 86 (2001) 5620-5623.

\bibitem{YoshizawaEL} A. Yoshizawa, R. Kaji, H.
Tsuchida, Electron. Lett. 39 (2003) 621-622.

\bibitem{YoshizawaJJAP} A. Yoshizawa, R. Kaji, H. Tsuchida,
Jpn. J. Appl. Phys. 42 (2003) 5652-5653.

\bibitem{Waks} E. Waks, E. Diamanti, B.C. Sanders, S.D. Bartlett, Y.
Yamamoto, Phys. Rev. Lett. 92 (2004) 113602/1-4.

\bibitem{Miller} A.J. Miller, S.W. Nam, J.M. Martinis, A.V. Sergienko,
Appl. Phys. Lett. 83 (2003) 791-793.

\bibitem{NamekataOL} N. Namekata, Y. Makino, S. Inoue, Opt. Lett.
27 (2002) 954-956.

\bibitem{Yuen} H.P. Yuen, Phys. Rev. A 13 (1976) 2226-2243.

\bibitem{Scully} M.O. Scully, M.S. Zubairy, Quantum Optics,
Cambridge University Press, Cambridge, 1997, Chap. 16.

\bibitem{Mollow} B.R. Mollow, R.J. Glauber, Phys. Rev. 160 (1967) 1076-1096.

\bibitem{deRiedmatten} H. de Riedmatten, V. Scarani, I. Marcikic, A. Ac\'{\i}n, W. Tittel, H. Zbinden, N.
Gisin, J. Mod. Opt. 51 (2004) 1637-1649.

\bibitem{Mori} S. Mori, N. Namekata, Y. Takamura, S. Inoue,
Generation of correlated photon-pair in the telecom wavelength,
in: Proceedings of the Seventh Quantum Information Technology
Symposium, IEICE, Tokyo, 2002, pp. 145-148 (in Japanese).

\bibitem{YoshizawaDischarge} A. Yoshizawa, R. Kaji, H. Tsuchida, Jpn. J. Appl. Phys.
43 (2004) L735-L737.

\bibitem{Banaszek} K. Banaszek, A.B. U'Ren, I.A. Walmsley, Opt. Lett. 26 (2001) 1367-1369.

\bibitem{NamekataPreparation} N. Namekata, S. Mori, S. Inoue, submitted.

\bibitem{Mason} E.J. Mason, M.A. Albota, F. K\"{o}nig, F.N.C. Wong, Opt. Lett. 27 (2002) 2115-2117.

\bibitem{Pelton} M. Pelton, P. Marsden, D. Ljunggren, M. Tengner, A. Karlsson, A. Fragemann, C.
Canalias, F. Laurell, Opt. Express 12 (2004) 3573-3580.

\bibitem{Fasel} S. Fasel, O. Alibart, S. Tanzilli, P. Baldi, A.
Beveratos, N. Gisin, H. Zbinden, New J. Phys. 6 (2004) 163/1-11.

\end{thebibliography}
\end{document}